\documentclass[12pt, amsmath]{revtex4-1}
\usepackage{graphicx}
\usepackage{amsmath,amssymb,mathrsfs,amsthm}
\usepackage[colorlinks]{hyperref}
\usepackage{cleveref}
\usepackage[dvipsnames]{xcolor}
\usepackage{ulem} 
\usepackage{float}
\usepackage{mathtools}
\usepackage{euscript}
\hypersetup{
    colorlinks=false,
    pdfborder={0 0 0},
}

\input{CQGformat.input}

\newcommand{\Zl}{\EuScript{Z}^\ell}
\newcommand{\Xl}{\EuScript{X}^\ell}

\newcommand{\pri}[1]{#1^{\prime}}
\newcommand{\ppri}[1]{#1^{\prime\prime}}

\def\Ener{\mathrm{e}}
\def\Pres{\mathrm{p}}
\def\Heat{\mathcal{Q}}

\def\Proj{\ensuremath{\perp}}

\def\TP{\ensuremath{\EuScript{T}}}

\def\Odd{\ensuremath{\EuScript{O}}}
\def\Even{\ensuremath{\EuScript{E}}}
\def\BB{\ensuremath{\EuScript{B}}}
\def\CC{\ensuremath{\EuScript{C}}}

\begin{document}

\title{Perturbations of relativistic dissipative stars}

\author{Jaime Redondo--Yuste}

\address{Center of Gravity, Niels Bohr Institute, Blegdamsvej 17, 2100 Copenhagen, Denmark}
\address{jaime.redondo.yuste@nbi.ku.dk}
\vspace{6pt}

\begin{abstract}

Viscous fluids can dissipate and alter the propagation of gravitational waves, as well as modify the relaxation and stability properties of self-gravitating fluids. This is particularly relevant in order to understand the relaxation to equilibrium of neutron stars, and their gravitational wave emission. Here we study the linearized theory of perturbations of spherically symmetric self-gravitating fluids, treating only the radiative modes. Dissipative effects are included through the hydrodynamics theory of Bemfica, Disconzi, Noronha and Kovtun (BDNK). This theory has been shown to be causal and stable, despite involving only first order gradients. We show how the problem reduces to two coupled wave equations in the axial sector, one of them associated to a novel viscous mode, and including explicitly dissipative terms. In the polar sector we reduce the problem to five coupled wave equations and one additional constraint. We comment on their causal structure, and recover the causality bounds of the BDNK theory. 
\end{abstract}


\section{Introduction}

Light interacts with matter in various ways: it can get absorbed or reflected, it can diffract or refract at interfaces, depending on the microscopic structure of the matter with which it interacts. This behavior is a double edge sword: on the one hand, knowledge of these interactions allows us to learn much about matter itself, just by studying how does light scatter off it. On the other hand, signals often get damped, dispersed, or affected in various ways due to their interaction with, e.g., the atmosphere, limiting our technological capabilities to perform precision science. 

Gravitational waves (GWs) also interact with matter~\cite{grishchuk1980gravitational}, albeit much more weakly, due to the smallness of Newton's constant. Because of its weak character, the interaction between GWs and matter is often very simple. For instance, a GW scattering off a perfect fluid results, simply, in a phase shift, that accounts for the redshift that the GW suffers when crossing the gravitational potential of the matter content~\cite{Cardoso:2022whc,Speeney:2024mas}. The situation is different in the presence of viscosity. Early work by Hawking~\cite{Hawking:1966qi} already showed that GWs propagating in a viscous, expanding universe, are damped. Several works studied this effect~\cite{esposito1971absorption,Weinberg:1971mx, Madore1973, szekeres1971linearized}, concluding that a medium with shear viscosity $\eta$ absorbs GWs on a timescale $c^2 G^{-1}\eta^{-1}$, where $c$ is the speed of light, and $G$ is Newton's constant. 

A remarkable work by Press~\cite{Press:1979rd} then conjectured that a highly viscous material could act as a mirror of GWs, or similarly, as a conductor. He dubbed such material ``Respondium''. However, he also concluded that for such a material to exist, and not undergo gravitational collapse (based on the hoop conjecture), it would violate the dominant energy condition. A clear limitation of the analysis is that it was based on Newtonian hydrodynamics on a weakly perturbed flat space. However, it was also hinted that in order to have a strong interaction between GWs and matter, spacetime curvature could become relevant, too~\cite{Press:1979rd}. A first attempt to work out a consistent description was carried out in~\cite{papadopoulos1985absorption}, although still in the context of an expansion in powers of $c$. It remains an open question whether a ``Respondium'' type material can exist. 

Understanding the interaction of GWs and viscous matter also becomes important in the context of precision GW astrophysics. GW interferometers have and will observe mergers of neutron stars (NSs), which then might relax to equilibrium, or collapse promptly to form a black hole. Viscous effects can affect their GW signal during the inspiral stage~\cite{HegadeKR:2023glb, Ripley:2023lsq, Ripley:2023qxo, Chia:2024bwc}, but also the GW emission during their final relaxation. Numerical simulations hint towards a significant impact, in particular, of bulk viscosity, in the late time relaxation of NSs~\cite{Alford:2010gw, Alford:2017rxf, Alford:2018lhf, Most:2021zvc, Most:2022yhe, Camelio:2022fds, Camelio:2022ljs, Chabanov:2023abq, Chabanov:2023blf}. This happens because newly formed NSs, such as super and hypermassive NSs formed after the merger of two stars, are very hot, and bulk viscosity becomes the dominant dissipative effect at high temperatures. However, the only perturbative estimates on the relaxation timescales due to dissipative effects~\cite{cutler1987effect, 1990ApJ...363..603C} are, indeed, estimates, based on the energy functional of~\cite{Lindblom:1983ps}. A first-principles calculation of the impact of different kinds of viscosity on the quasinormal modes of hot NSs, their precise damping times, and so on, is still lacking. 

Our main motivation is to set up the stage to fill this gap in the literature. We will do so by studying, in a first-principles manner, the oscillations of self-gravitating, dissipative fluids. We will restrict ourselves to spherical symmetry for simplicity. In this context, our goal is to derive the equations of motion for the coupled matter and gravitational perturbations in the interior of the fluid, for those modes that radiate GWs. In this work, we will provide evolution equations for axial and polar parity gravitational and fluid perturbations, ready to be studied numerically for particular stellar models. 

In order to do this, we need to specify a particular theory of dissipative hydrodynamics. The naive extensions of the Navier--Stokes hydrodynamics to the relativistic regime, first undertaken by Eckart~\cite{Eckart:1940te} and Landau~\cite{landau2013fluid} are problematic: they lead to instabilities because of their acausal character~\cite{Hiscock:1983zz, Hiscock:1985zz, Detweiler:1985zz}. There is a way to heal these pathologies, by extending the theory and including effects which are second order in the gradients. This is sometimes referred to as the Israel--Stewart theory~\cite{Israel:1979wp, Rezzolla:2013dea}, or as extended irreversible thermodynamics. However, the inclusion of second order effects complicates significantly the analysis. Recent work tackled the problem of linearized perturbations in this theory~\cite{Diaz-Guerra:2024gff}. An alternative approach, based on a variational framework, was put forward by Carter~\cite{Carter:1985abc, Carter:1988def, Carter:2006ghi, Andersson:2020phh}. Perhaps the most natural way to solve the problems that the Eckart or Landau theories suffer from is solved simply by including all possible first order corrections to the stress energy tensor. This means including also timelike gradients, and not just spacelike ones. By following this line of thought, Bemfica, Disconzi, Noronha and Kovtun (BDNK)~\cite{Bemfica:2017wps, Bemfica:2019cop, Bemfica:2019knx, Bemfica:2020zjp, Kovtun:2019hdm} showed that one can obtain a theory that is (i) causal, provided the transport coefficients satisfy some constraints, (ii) linearly stable, and (iii) the stress energy tensor only involves up to first order covariant derivatives of the thermodynamical variables. For this reason, in this work we will consider the stress energy tensor of a fluid in the BDNK theory as a starting point. 

After specifying the stress energy tensor, we need to deal with perturbations of a non-vacuum solution. We exploit the spherical symmetry of the background, decomposing all the tensors in a $2+2$ expansion, following work by Gerlach and Sengupta~\cite{Gerlach:1979rw, Gerlach:1979ze, Gerlach:1980tx}. Additionally, since the fluid background velocity induces a preferred frame, we follow work by Gundlach and Mart\'in-Garc\'ia~\cite{Gundlach:1999bt, Martin-Garcia:1998zqj, Martin-Garcia:2000cgm} to project the linearization of the Einstein equations into scalars. This formalism also allows us to work in a straightforward way in terms of gauge invariant quantities. 

Linear perturbations decompose according to its parity in axial (odd), and polar (even). In the axial sector, we reduce the system to two coupled wave equations. One of them describes the propagation of GWs inside the star, and includes explicit dissipation effects due to the shear viscosity. The second equation corresponds to a viscous mode, with no analogue in the perfect fluid case. This is akin to the appearance of a second sound mode for superfluid stars~\cite{1994ApJ...421..689L,Comer:1999rs, Andersson:2001bz}, or with the appearance of an unstable pair of modes in classical hydrodynamics for the Poiseuille flow, which are absent in the inviscid case~\cite{heisenberg1985stabilitat, 1985JFM...151..189L}. This mode corresponds to a rotational motion in the star, which oscillates with a frequency proportional to the shear viscosity. In the polar sector, the situation is more complex, and we find five coupled wave equations, together with a constraint. We provide a \texttt{Wolfram} language package and scripts ready to use~\cite{Redondo-Yuste:2024GITHUB}, including the derivation of these equations. Additionally, we study their causal structure. We find that in the polar sector there are propagating modes associated to GWs, as well as a viscous mode, very similar to the axial sector. Additionally, there is a last pair of modes, corresponding to fluid density perturbations and bulk viscosity driven dynamics, respectively. Despite obtaining them from the linearized theory with respect to a background which is not flat, the causal structure of our equations recovers immediately the BDNK causality constraints of~\cite{Bemfica:2020zjp}. 

The structure of the paper is the following. In Section~\ref{sec:GSGM} we review the gravitational perturbations  in the $2+2$ formalism, projected onto a particular causal frame. Section~\ref{sec:BDNK} reviews BDNK hydrodynamics, and computes the relevant projections of the perturbed stress energy tensor, in terms of the fluid perturbations. In Section~\ref{sec:Coordinates} we briefly summarize how to change from the coordinate--free language employed up to this point, to the particular set of coordinates that we use to derive the master equations. Finally, Sections~\ref{sec:Odd} and~\ref{sec:Even} show the derivation of the reduced system of odd and even parity wave equations, respectively. We summarize our findings, and comment on novel research directions that build upon this work in Section~\ref{sec:Conclusions}.

Unless otherwise specified, we use geometric units with $G=c=1$. Upper case latin indices $A,B,\dots$ are indices on a two dimensional Lorentzian manifold. Lower case latin indices $a,b,\dots$ are indices on the sphere. Greek indices $\mu,\nu,\dots$ are spacetime indices. Our notation is summarized in Table~\ref{tab:Notation}.

\begin{table}[]
\begin{tabular}{ccc}
\hline
Symbol(s) & Definition & First appears \\ \hline
$\{l_A,n_A\}$ & Basis on $\mathcal{M}^2$ & ~\eqref{eq:2D_Metric} \\
$\{g_{AB},p_{AB},q_{AB},\epsilon_{AB}\}$ & Basis of symmetric, rank 2 tensors on $\mathcal{M}^2$ & \eqref{eq:2D_Metric}--\eqref{eq:2D_Projectors} \\
$\{U,W,\mu,\nu\}$ & Background scalars & \eqref{eq:Background_Scalars_Def} \\
$\lambda^2$ & $\ell(\ell+1)$ & \eqref{eq:Odd_EEqs_Projected} \\
$\{h_p,h_q,h_g,h_S\}$ & Even parity metric perturbations & ~\eqref{eq:Metric_SET_Decomposition} \\
$\{k_n,k_l\}$ & Odd parity metric perturbations & ~\eqref{eq:Metric_SET_Decomposition} \\
$\{\TP_p,\TP_q,\TP_g,\TP_l,\TP_n,\TP_S,\Tilde{\TP}_S\}$ & Even parity stress energy tensor perturbations & ~\eqref{eq:Metric_SET_Decomposition} \\
$\{\vartheta_n,\vartheta_l,\vartheta_S\}$ & Odd parity stress energy tensor perturbations & ~\eqref{eq:Metric_SET_Decomposition} \\
$\{\Even_p,\Even_q,\Even_g,\Even_l,\Even_n,\Even_S,\Tilde{\Even}_S\}$ & Even parity linearized Einstein tensor & ~\eqref{eq:Einstein_Decomposition} \\
$\{\Odd_n,\Odd_l,\Odd_S\}$ & Odd parity linearized Einstein tensor & ~\eqref{eq:Einstein_Decomposition} \\
$\{\BB_n,\BB_l,\BB_S,\CC_S\}$ & Linearized conservation of stress energy tensor & ~\eqref{eq:Conservation_SET_Decomposition} \\
$\Ener$, $\Pres$ & Background energy density and pressure & ~\eqref{eq:Constitutive_Relations} \\
$\rho$ & $\Ener+\Pres$ & ~\eqref{eq:Constitutive_Relations} \\
$c_s$ & $\sqrt{d\Pres / d\Ener}$ (Sound speed) & ~\eqref{eq:Constitutive_Relations} \\
$\{\zeta,\eta,\tau_\Ener,\tau_\Pres,\tau_\Heat\}$ & Transport coefficients & ~\eqref{eq:Constitutive_Relations} \\
$\{V,\tau_\pm, V_\pm\}$ & Transport coefficient combination shorthands & ~\eqref{eq:Transport_Coefs_Shorthands} \\
$\{\alpha,\beta,\gamma,\omega\}$ & Fluid velocity and energy density perturbations & ~\eqref{eq:Perturbations_Fluid} \\
$H$ & $c_s^2\Ener\omega/\rho$ & ~\eqref{eq:Enthalpy_Definition} \\
$\{\Phi,\Lambda, M\}$ & Metric coordinate functions & ~\eqref{eq:Metric_In_Coordinates} \\
$\psi$ & $r^{-1}e^{\Phi/2}k_n$ & ~\eqref{eq:Master_Odd_Inviscid} \\\hline 
\end{tabular}
\caption{\label{tab:Notation} Summary of the main variables used in this paper, including their brief definition, and the equation where they first appear.}
\end{table}

\section{GSGM Formalism for spacetime perturbations}\label{sec:GSGM}

We use spherical symmetry to split the spacetime manifold as a warped product of a sphere and a two dimensional Lorentzian manifold, i.e., $\mathcal{M}=\mathcal{M}^2\times_{r^2}\mathcal{S}^2$. This decomposition allows us to reduce Einstein equations, and its perturbations, to equations on $\mathcal{M}^2$, by exploiting the spherical symmetry of the background. Gerlach and Sengupta first used this idea to obtain a particularly elegant derivation of the equations describing linearized perturbations around a Schwarzschild black hole~\cite{Gerlach:1979rw, Gerlach:1980tx}. Since then, this formalism has been later revisited and adapted to different set-ups, ranging from higher dimensions~\cite{Kodama:2000fa, Ishibashi:2003ap, Kodama:2003jz, Kodama:2003kk}, further restrictions to vacuum spacetimes~\cite{Martel:2005ir}, including deriving the equations from the curvature wave equation~\cite{Mukkamala:2024dxf}, higher order perturbations~\cite{Brizuela:2006ne, Brizuela:2007zza}, perturbations in modified gravity theories~\cite{Ripley:2017kqg, Tattersall:2017erk}, gauge theories~\cite{Pereniguez:2023wxf}, and non-stationary spacetimes~\cite{Redondo-Yuste:2023ipg}. 

A notable advancement was introduced by Gundlach and Mart\'in--Garc\'ia~\cite{Gundlach:1999bt, Martin-Garcia:1998zqj, Martin-Garcia:2000cgm} (hence we dub this formalism GSGM, following~\cite{Passamonti:2005ac}). It relies on introducing a frame $\{l_A, n_A\}$ on $\mathcal{M}^2$, where $l^Al_A=-1$, $n^An_A=1$ and $l^An_A=0$. This allows to project the equations onto the frame directions, reducing them to scalar equations. This framework is particularly useful when studying perturbations where the matter sector is described by a fluid, since then the fluid velocity is a natural choice for a frame vector, that fixes completely the frame. A remarkable achievement of this formalism was the characterization of the coupling between radial and non--radial modes, discussed in detail in~\cite{Passamonti:2004je, Passamonti:2005ac}. 

\subsection{Background geometry}

First, we consider spherically symmetric solutions to Einstein's equations with an arbitrary stress--energy tensor. We choose coordinates $\{y^A\}$ on $\mathcal{M}^2$, and coordinates $\{\theta^a\}$ on $\mathcal{S}^2$. From now on, upper case (respectively lower case) latin indices are indices on $\mathcal{M}^2$ (respectively $\mathcal{S}^2$). The metric is written as 
\begin{equation}\label{eq:Background_Metric_Decomposition}
    ds^2 = g_{AB}(y)dy^A dy^B + r^2(y)\gamma_{ab}d\theta^a d\theta^b \, , 
\end{equation}
where $r$ is the warping factor (areal radius) and $\gamma_{ab}$ is the usual metric on the $2$--sphere. Let us introduce $v_A = \nabla_A r$. Here, $\nabla$ denotes covariant derivatives on $\mathcal{M}^2$ compatible with the metric $g_{AB}$, and $D$ the covariant derivative on the sphere. 

Using the frame $\{l^A,n_A\}$ we can alternatively write the metric as 
\begin{equation}\label{eq:2D_Metric}
    g_{AB} = -l_Al_B+n_An_B \, .
\end{equation}
Additionally, there are only three other independent rank--2 tensors, which we label
\begin{equation}\label{eq:2D_Projectors}
    p_{AB} = l_Al_B+n_An_B \, , \quad q_{AB} = l_An_B+n_Al_B \, , \quad \epsilon_{AB}=-l_An_B+n_Al_B \, .
\end{equation}
A general stress energy tensor can then be written as 
\begin{equation}\label{eq:Background_SET_Decomposition}
    T_{\mu\nu}dx^\mu dx^\nu = \frac{1}{2}\Bigl(t_g g_{AB}+t_p p_{AB}-t_q q_{AB}\Bigr)dy^Ady^B + r^2 t_S\gamma_{ab}d\theta^a d\theta^b \, ,
\end{equation}
i.e., it is fully characterized by the four scalars $\{t_g, t_p, t_q, t_S\}$. The normalization of the previous expression is chosen so that, e.g., $q^{AB}T_{AB}=t_q$. 

We further introduce the frame derivatives
\begin{equation}\label{eq:Frame_Derivatives_Def}
    \dot{f} = l^A\nabla_A f \,  \qquad \pri{f}=n^A\nabla_A f \, ,
\end{equation}
and the scalars
\begin{equation}\label{eq:Background_Scalars_Def}
    U = l^Av_A \, , \quad W = n_Av^A \, , \quad \mu = \nabla_Al^A \, , \quad \nu = \nabla^An_A \, .
\end{equation}
Straightforward computations yield the following useful identities, which we will apply repeatedly in what follows
\begin{equation}\label{eq:Frame_Identities}
    \begin{aligned}
        \nabla_A l_B =& n_B(\mu n_A - \nu l_A) \, , \qquad
        \nabla_A n_B = u_B (\mu n_A - \nu l_A) \, , \\[4pt]
        v^Av_A =& -U^2+W^2 \, , \qquad
        (\dot{f})^\prime-\dot{(f^\prime)} = \mu f^\prime - \nu \dot{f} \, .
    \end{aligned}
\end{equation}
We now project Einstein equations onto the frame, obtaining
\begin{equation}\label{eq:Background_Einstein_General}
    \begin{aligned}
        W^\prime-\dot{U}+\nu W - \mu U + 2v^2 - r^{-2} =& 4\pi t_g \, , \\[4pt]
        -W^\prime-\dot{U}+\nu W + \mu U + v^2 - 2 W^2 =& 4\pi t_p \, , \\[4pt]
        -\dot{W}-U^\prime+\mu W + \nu U - 2UW =& 4\pi t_q \, , \\[4pt]
        -\dot{U}+W^\prime+\nu W-\mu U+v^2 - \frac{1}{2}R^{(2)} =& 8\pi t_S \, .
    \end{aligned}
\end{equation}
We can use these equations to eliminate three of the first order frame derivatives of $U,W$. The only frame derivative that cannot be eliminated from these equations can be instead obtain from the other $3$ via commutation relations~\eqref{eq:Frame_Identities}:
\begin{equation}\label{eq:UPrime_Background_Equation}
    U^\prime = (\mu - U)W - 2\pi t_q \, .
\end{equation}
Finally, recall that for a two dimensional manifold, $R^{(2)}g_{AB} =2R_{AB}$, so we can eliminate the Ricci scalar in terms of the scalars defined in~\eqref{eq:Background_Scalars_Def} as 
\begin{equation}
    \frac{1}{2}R^{(2)} = \dot{\mu}-\pri{\nu}+\mu^2-\nu^2 = -\dot{U}+W^\prime+\nu W-\mu U+v^2-8
    \pi t_S\, .
\end{equation}
We will use the last equality to eliminate every instance of $\dot{\mu}-\pri{\nu}$. 

\subsection{Linear perturbations}

We now consider linearized perturbations of the metric and the stress--energy tensor. We expand them in tensor spherical harmonics following~\cite{Martin-Garcia:1998zqj, Martin-Garcia:2000cgm}. The tensor spherical harmonics $\{\Zl, \Xl\}$ are introduced in the Supplemental Material (SM). We also write already the metric perturbation in terms of gauge invariant variables (equivalently, in the Regge-Wheeler gauge), and discuss gauge invariance in the SM. 
In what follows, we consider only radiative modes, i.e., modes with spherical harmonic number $l \geq 2$. Treating the non-radiative monopolar and dipolar modes is left for future work --the expansion below in tensor spherical harmonics simplifies for $l=0,1$ modes (e.g. $\Xl_a=\Xl_{ab}=0$ for spherically symmetric $l = 0$ modes), and residual gauge transformations can be used to further simplify their treatment. These non radiative modes do not excite propagating gravitational waves, but have important implications for the linear stability of stellar configurations.

The metric and stress energy tensor perturbations can be written as
\begin{equation}\label{eq:Metric_SET_Decomposition}
    \begin{aligned}
        g^{(1)}_{\mu\nu}dx^\mu dx^\nu =& \frac{1}{2}\Bigl(h^{\ell}_g g_{AB}+ h^{\ell}_p p_{AB} -  h^{\ell}_q q_{AB}\Bigr)\Zl dy^Ady^B \\[4pt]
        &+2\Bigl( k^{\ell}_n n_A- k^{\ell}_l l_A\Bigr)\Xl_a dy^A d\theta^a +r^2 h^{\ell}_S \Zl \gamma_{ab}d\theta^a d\theta^b \, , \\[6pt]
        T^{(1)}_{\mu\nu}dx^\mu dx^\nu =& \frac{1}{2}\Bigl( \TP^{\ell}_g g_{AB}+ \TP^{\ell}_p p_{AB} -  \TP^{\ell}_qq_{AB}\Bigr)\Zl dy^Ady^B \\[4pt]
        &+2\Bigl[\Bigl( \vartheta^{\ell}_n n_A - \vartheta^{\ell}_l l_A\Bigr)\Xl_a + \Bigl( \TP^{\ell}_n n_A - \TP^{\ell}_l l_A\Bigr)\Zl_a\Bigr]dy^Ad\theta^a \\[4pt]
        &+\Bigl[\vartheta^{\ell}_S \Xl_{ab}+r^2\Bigl(\TP^{\ell}_S\Zl\gamma_{ab} + \Tilde{\TP}^{\ell}_S\Zl_{ab}\Bigr)\Bigr]d\theta^a d\theta^b \, .
    \end{aligned}
\end{equation}
Even parity perturbations are described by the $\{h, \TP\}$ variables, while odd parity perturbations are described by the $\{k,\vartheta\}$. The index $\ell$ denotes the associated spherical harmonic pair of indices $\ell = (l,m)$, and will be omitted from now on, assuming we work with a single harmonic mode. From these metric perturbations we can now obtain the $10$ projections of Einstein equations. We can decompose the Einstein tensor in a similar manner as the stress energy tensor, 
\begin{equation}\label{eq:Einstein_Decomposition}
    \begin{aligned}
        G_{\mu\nu}dx^\mu dx^\nu =& \frac{1}{2}\Bigl(\Even_g g_{AB}+\Even_p p_{AB} -\Even_q q_{AB}\Bigr)\Zl dy^Ady^B \\[4pt]
        &+2\Bigl[\Bigl(\Odd_n n_A - \Odd_l l_A\Bigr)\Xl_a + \Bigl(\Even_n n_A - \Even_l l_A\Bigr)\Zl_a\Bigr]dy^Ad\theta^a \\[4pt]
        &+\Bigl[\Odd_S\Xl_{ab}+r^2\Bigl(\Even_S \Zl\gamma_{ab} + \Tilde{\Even}_S\Zl_{ab}\Bigr)\Bigr]d\theta^a d\theta^b \, ,
    \end{aligned}
\end{equation}
where $\{\Odd,\Even\}$ denote odd and even parity, respectively. 

Additionally we have the linearized perturbation of the conservation of the stress energy tensor $\Delta[\nabla_\mu T^\mu_\nu]$, is given by 
\begin{equation}\label{eq:Conservation_SET_Decomposition}
    \Delta[\nabla_\mu T^\mu_\nu]dx^\nu = \Bigl(\BB_n n_A - \BB_l l_A\Bigr)\Zl dy^A + \Bigl(\BB_S \Zl_a + \CC_S \Xl_a\Bigr)d\theta^a \, ,
\end{equation}
where $\{\BB,\CC\}$ are even and odd equations, respectively, which we write in the SM. We write below the projections of Einstein equations~\footnote{Remember that Einstein's equations imply, through the Bianchi identities, the conservation of the stress energy tensor.}.

The projected Einstein equations in the odd sector are given by 
\begin{equation}\label{eq:Odd_EEqs_Projected}
    \begin{aligned}
        \Odd_l \equiv& \dot{\pri{k}}_n-\ppri{k}_l +2(\mu-U)\pri{k}_n+(2W-\nu)\dot{k}_n-\nu\pri{k}_l+\Bigl(4\pi t_q +\pri{\mu}-2UW\Bigr)k_n \\
        &+ \Bigl(4\pi(t_g-t_p+4t_S)+2(W^2 + \mu U-\nu W)-v^2-\pri{\nu}+\frac{\lambda^2-1}{r^2}\Bigr)k_l = 16\pi\vartheta_l \, , \\[6pt]
        \Odd_n \equiv& \ddot{k}_n-\dot{\pri{k}}_l+\mu\dot{k}_n-2U\pri{k}_l+(2W-\nu)\dot{k}_l+\Bigl(2UW-4\pi t_q-\pri{\nu}\Bigr)k_l\\
        &+\Bigl(4\pi(2t_g+t_p+2t_S)+2(\mu U-\nu W-W^2)+\pri{\nu}+\nu^2-\mu^2+\frac{\lambda^2}{r}\Bigr)k_n = 16\pi\vartheta_n \, ,\\[6pt]
        \Odd_S \equiv &  \pri{k}_n+\nu k_n-\dot{k}_l-\mu k_l = 8\pi \vartheta_S \, . 
    \end{aligned}
\end{equation}
To avoid confusion, whenever we write a double derivative we do it in the order $\dot{\pri{f}} = l^A\nabla_A\Bigl(n_B\nabla^B f\Bigr)$ (this is, the prime derivative acts before the dot derivative). The even parity equations were first written in~\cite{Martin-Garcia:1998zqj}. We recover them here, where the relation between our notation and theirs (MG) is $2\eta^{MG} \mapsto h_g$, $2\phi^{MG}\mapsto h_p$, $2\psi^{MG}\mapsto -h_q$, $k^{MG}\mapsto h_S$. For convenience, we write them below:
\begin{equation}\label{Even_EEqs_Projected}
    \begin{aligned}
        \Even_g \equiv&  -\ddot{h}_S+\ppri{h}_S  +U(\pri{h}_q-\dot{h}_p)+W(\dot{h}_q-\pri{h}_p)+(\nu+4W)\pri{h}_S-(\mu+4U)\dot{h}_S \\
        &+h_p\Bigl(8\pi t_p+v^2-2\mu U-2W(\nu+W)\Bigr)+2h_q\Bigl(\nu U+(\mu+U)W-4\pi \TP_q\Bigr)\\
        &-\frac{1}{2r^2}\Bigl((\lambda^2+2)h_g+2(\lambda^2-2)h_S\Bigr) = 8\pi t_g \, , \\[6pt]
        \Even_p \equiv& -\ddot{h}_S-\ppri{h}_S+U(\dot{h}_g-\pri{h}_q)+W(\dot{h}_q+\pri{h}_g)+(\mu-2U)\dot{h}_S+(\nu-2W)\pri{h}_S \\
        &+\Bigl(8\pi t_g+2\mu U-2\nu W-v^2+\frac{\lambda^2+2}{2r^2}\Bigr)h_p = 8\pi \TP_p \, , \\[6pt]
        \Even_q \equiv& -2\dot{\pri{h}}_S+U(\pri{h}_g-\pri{h}_p)+W(\dot{h}_g+\dot{h}_p)+2\Bigl[(\nu-W)\dot{h}_S-\pri{h}_S\Bigr]\\
        &+\Bigl(8\pi t_G+2\mu U-2\nu W-v^2+\frac{\lambda^2+2}{2r^2}\Bigr)h_q = 8\pi \TP_q \, , \\[6pt]
        \Even_l \equiv& \pri{h}_q-\dot{h}_g-\dot{h}_p-2\dot{h}_S+2\Bigl(Uh_g+\nu h_q-\mu h_P) = 32\pi\TP_l \, , \\[6pt]
        \Even_n \equiv& \pri{h}_p-\pri{h}_g-\dot{h}_q-2\pri{h}_S+2\Bigl(Wh_g+\nu h_p-\mu h_q\Bigr) = 32\pi\TP_n \, , \\[6pt]
        \Tilde{\Even}_S \equiv& h_g = -16\pi r^2\Tilde{\TP}_S \, , \\[6pt]
        \Even_S \equiv& 2\ppri{h}_S-\ppri{h}_p+\ppri{h}_g-2\ddot{h}_S-\ddot{h}_p-\ddot{h}_g+2\dot{\pri{h}}_q+2\Bigl[(\nu+2W)\pri{h}_S-(\mu+2U)\dot{h}_S\Bigr]\\
        &-(3\mu+2U)\dot{h}_p-(3\nu+2W)\pri{h}_p-\mu\dot{h}_g+\nu\pri{h}_g+2\Bigl[(2\mu+U)\pri{h}_q+(\nu+W)\dot{h}_q\Bigr]\\
        &+2\Bigl[4\pi(2t_S+t_P-t_G)+v^2-2\mu U-2\nu W-2\nu^2-2\pri{\nu}-r^{-2}\Bigr]h_p\\
        &+2\Bigl[2\nu(\mu+U)+2\mu W+\dot{\nu}-\pri{\mu}-4\pi t_q\Bigr]h_q-\Bigl(16\pi t_S+\frac{\lambda^2}{r^2}\Bigr)h_g + 32\pi t_Sh_S = 32\pi \TP_S \, , 
    \end{aligned}
\end{equation}
where $\lambda^2=l(l+1)$. 

\section{First order relativistic hydrodynamics}\label{sec:BDNK}
Until this point the stress energy tensor was left unspecified. In this section, we will specify it to be the stress energy tensor of an imperfect fluid. We refer the interested reader to~\cite{Bemfica:2020zjp, Rocha:2023ilf} for a historical overview of the topic of relativistic, dissipative fluid dynamics, including a discussion of the BDNK theory, as well as other approaches. 

In this section, we first discuss some generalities about the BDNK theory. Then, we compute the frame projections of the background stress energy tensor. Finally, we introduce the perturbations of the fluid degrees of freedom, and compute the perturbations of the scalars entering the equations in the previous section.

\subsection{The general theory}

The most general stress energy tensor for a fluid is
\begin{equation}\label{eq:BDNK_SET}
    T_{\mu\nu} = \mathcal{E}u_\mu u_\nu + \mathcal{P}\perp_{\mu\nu} + u_{\mu}\Heat_{\nu} + \Heat_{\mu}u_{\nu}+\mathcal{T}_{\mu\nu} \, ,
\end{equation}
where $u_\mu$ is the fluid velocity, and $\perp_{\mu\nu} = g_{\mu\nu}+u_\mu u_\nu$. Above, the heat current $\mathcal{Q}_\mu$ is transverse to the velocity, $u^\mu\mathcal{Q}_\mu = 0$, and the tensor $\mathcal{T}_{\mu\nu}$ is both traceless and transverse, $\mathcal{T}_{\mu\nu}u^\nu = 0$, $\mathcal{T}_\mu^\mu = 0$.

The quantities $\mathcal{E},\mathcal{P},\mathcal{Q},\mathcal{T}$ need to be specified as functions of some thermodynamic variables. A general fluid is described by three variables: the fluid velocity $u^\mu$, its energy density $\Ener$ and its particle number density. In this work, in order to simplify the problem, we will restrict to barotropic matter. In particular, this corresponds to setting the heat conductivity in Ref.~\cite{Bemfica:2020zjp} to zero, and neglecting the effects of particle and temperature diffusion. The heat current, however, will not vanish, as necessary to obtain a causal and stable dissipative theory. We leave for future explorations the inclusion of heat diffusion, and a chemical potential.

Now, specifying how do the terms of the stress energy tensor depend on the thermodynamic variables $\{u^\mu,\Ener\}$ characterizes the fluid theory. For a perfect fluid, for example, one just requires that $\mathcal{E}=\Ener$, $\mathcal{P}=\Pres$ is the pressure, which is related to the energy by some equation of state, and $\mathcal{Q}=\mathcal{T}=0$. For BDNK fluids we consider additionally terms involving first order spacetime derivatives of the thermodynamic variables. In detail, the constitutive relations are~\cite{Bemfica:2020zjp}
\begin{equation}\label{eq:Constitutive_Relations}
    \begin{aligned}
        \mathcal{E} =& \Ener + \tau_{\Ener}\Bigl[u^\mu\nabla_\mu\Ener + \rho\nabla_\mu u^\mu\Bigr] \, , \qquad 
        \mathcal{P} = \Pres - \zeta \nabla_\mu u^\mu + \tau_{\Pres} \Bigl[u^\mu\nabla_\mu\Ener + \rho\nabla_\mu u^\mu\Bigr] \, , \\
        \Heat_\mu =& \tau_\Heat\Bigl[\rho u^\nu\nabla_\nu u^\mu +c_s^2\Proj_{\mu\nu}\nabla^\nu \Ener\Bigr]  \, , \qquad 
        \mathcal{T}_{\mu\nu} = -2\eta\sigma_{\mu\nu} \, ,
    \end{aligned}
\end{equation}
where we have introduced the shorthand $\rho = \Ener+\Pres$, $\sigma_{\mu\nu}$ is the shear associated to $u_\mu$, and $\{\tau_\Ener, \tau_\Pres, \tau_\Heat, \eta, \zeta\}$ are the transport coefficients. Notice that the transport coefficients $\tau_{\Ener,\Pres,\Heat}$ are effectively relaxation timescales, which are necessary to be added to make the theory causal. We have also introduced the sound speed $c_s^2 = d\Pres/d\Ener$. We will also introduce the following combinations of transport coefficients
\begin{equation}\label{eq:Transport_Coefs_Shorthands}
    V = \zeta + \frac{4}{3}\eta \, ,  \qquad \tau_{\pm} = \tau_\Ener \pm \tau_\Pres \, , \qquad V_\pm = V \pm \rho \tau_\mp \, ,
\end{equation}
in order to simplify some expressions. 

The transport coefficients are obtained from the microscopic theory. For example, bulk viscosity in NSs is directly related to neutrino transport~\cite{Camelio:2022fds, Camelio:2022ljs}. In order for the resulting hydrodynamic theory to be causal and stable, they must satisfy certain inequalities~\cite{Bemfica:2020zjp}. For example, in a companion work~\cite{Boyanov:2024jge} we choose a particular dependence of the transport coefficients on the stellar structure, which in turn constraints the maximum shear viscosity. However, in order to be as general as possible, here we just assume that they are smooth functions of the radial coordinate alone (in particular, we assume that they do not vary with time).

From these constitutive relations, we can immediately obtain the components of the stress energy tensor of the background. Spherical symmetry requires the fluid velocity to be $u_\mu dx^\mu = u_A dy^A$.
As is natural, we choose a frame where $l_A = u_A$, and the normal vector $n_A = \epsilon_{AB}u^B$ is specified in terms of the antisymmetric symbol $\epsilon_{AB}$ associated to the background metric $g_{AB}$. Then, the constitutive relations become  
\begin{equation}\label{eq:Constitutive_Relations_Projected}
    \begin{aligned}
        \mathcal{E} =& \Ener + \tau_\Ener\Bigl(\mu\rho+\dot{\Ener}\Bigr) \, , \qquad \qquad
        \mathcal{Q}_\mu dx^\mu = \tau_\Heat \Bigl(\nu\rho+c_s^2\pri{\Ener}\Bigr) n_Ady^A \, , \\
        \mathcal{P} =& \Pres -\zeta\mu + \tau_\Pres\Bigl(\mu\rho+\dot{\Ener}\Bigr) \, , \qquad
        \mathcal{T}_{\mu\nu}dx^\mu dx^\nu = \frac{2}{3}\eta\mu\Bigl(-2n_An_B dy^A dy^B+r^2\gamma_{ab}d\theta^a d\theta^b\Bigr) \, .
    \end{aligned}
\end{equation}
From these, it is straightforward to obtain the components of the stress energy tensor.
\begin{equation}\label{eq:Sources_Background}
    \begin{aligned}
        t_g =& \Pres-\Ener -\mu\Bigl(\zeta+\frac{4\eta}{3}\Bigr)-(\mu\rho+\dot{\Ener})(\tau_\Ener-\tau_\Pres) \, , \\
        t_p =& \Pres+\Ener -\mu\Bigl(\zeta+\frac{4\eta}{3}\Bigr)+(\mu\rho+\dot{\Ener})(\tau_\Ener+\tau_\Pres)\, , \\
        t_q =& -2\tau_\Heat\Bigl(\nu\rho+c_s^2\pri{\Ener}\Bigr) \, , \\
        t_S =& \Pres-\mu\Bigl(\zeta-\frac{2}{3}\eta\Bigr)+\tau_\Pres(\mu\rho+\dot{\Ener}) \,  .
    \end{aligned}
\end{equation}
One can already notice, at this level, that if the background is stationary, $\mu=0$, and the dot derivatives (which are proportional to derivatives with respect to the time coordinate) vanish. Hence, the viscous corrections drop immediately from $t_g$, $t_p$ and $t_S$. Spherical symmetry of the background will also guarantee that $t_q$ vanishes. 

\subsection{Fluid perturbations}

We consider now the perturbations to the fluid degrees of freedom. For simplicity we will keep the same frame as we used in the background fixed, i.e., $l_A = u^{(0)}_A$, where the zero index refers to the background value of the fluid velocity. 

The background energy density and pressure are labeled by $\Ener$ and $\Pres$. We will drop the $(0)$ index, since it will be understood that from now on whenever we write $\{u_A,\Ener,\Pres\}$ we refer to their background values.  Their perturbations can be expanded in the following way 
\begin{equation}\label{eq:Perturbations_Fluid}
    \begin{aligned}
        u^{(1)}_\mu dx^\mu =& \Bigl(\gamma n_A + \frac{1}{2}g^{(1)}_{AB}u^B\Bigr)\Zl dy^A + \Bigl(\alpha\Zl_a+\beta \Xl_a\Bigr)d\theta^a \, , \\
        \Ener^{(1)} =& \omega \Ener \Zl \, , \qquad \Pres^{(1)} = c_s^2\Ener^{(1)} \, .
    \end{aligned}
\end{equation}
Thus, odd parity perturbations are generated only by $\beta$, while even parity perturbations are described by $\{\alpha,\gamma,\omega\}$. We will also introduce the enthalpy perturbation $H$, which is related to $\omega$ by 
\begin{equation}\label{eq:Enthalpy_Definition}
    H = \frac{c_s^2\Ener\omega}{\Ener+\Pres} \, ,
\end{equation}
and will be useful later.

Now, in terms of these perturbed variables, we construct the stress energy tensor, and evaluate each of its components. To emphasize that we have now fixed the generic frame $\{l_A,n_A\}$ in terms of the background fluid velocity, to be $\{u_A,\epsilon_{AB}u^B\}$, we use the sub-index $u$ instead of $l$ to denote the contraction with the first frame vector. A straightforward but lengthy calculation yields
\begin{equation}\label{eq:Sources_Odd}
    \begin{aligned}
        \vartheta_u =& -\Ener k_u-\rho\beta-\tau_\Heat \rho \Bigl(\dot{\beta}+\dot{k}_u\Bigr) \, , \\
        \vartheta_n =& \Pres k_n - \Bigl[(\nu-2W)\beta+\pri{\beta}+\dot{k}_n\Bigr]\eta \, , \\
        \vartheta_S =& -2\beta\eta \, ,
    \end{aligned}
\end{equation}
for the odd sector, and 
\begin{equation}\label{eq:Sources_Even}
    \begin{aligned}
        \TP_g =& \frac{1}{2}\rho h_p + \frac{1}{2}(\Pres-\Ener)h_g-(1-c_s^2)\Ener\omega +(V_+-2\eta)\Biggl(
        \frac{1}{2}Wh_q-\dot{h}_S+\frac{\lambda^2\alpha}{r^2}-2W\gamma\Biggr)\\
        &+\frac{V_+}{4}\Bigl(\nu h_q-\dot{h}_g-\dot{h}_p+\pri{h}_q-4\pri{\gamma}-4\nu\gamma\Bigr)-\tau_-\Bigl[\Ener\dot{\omega}+\pri{\Ener}\Bigl(\gamma-\frac{h_q}{4}\Bigr)\Bigr] \, , \\
        \TP_p =& \frac{1}{2}\rho h_g + \frac{1}{2}(\Pres-\Ener)h_p + (1+c_s^2)\Ener\omega + (V_--2\eta)\Biggl(\frac{1}{2}Wh_q-\dot{h}_S+\frac{\lambda^2\alpha}{r^2}-2W\gamma\Biggr)\\
        &\frac{V_-}{4}\Bigl(\nu h_q-\dot{h}_g-\dot{h}_p+\pri{h}_q-4\pri{\gamma}-4\nu\gamma\Bigr)+\tau_+\Bigl[\Ener\dot{\omega}+\pri{\Ener}\Bigl(\gamma-\frac{h_q}{4}\Bigr)\Bigr] \, , \\
        \TP_q =& \frac{1}{2}(\Pres-\Ener)h_q-2\rho\gamma - \frac{\tau_\Heat}{2}\Biggl[  \rho\Bigl(\dot{h}_q+4\dot{\gamma}+\pri{h}_g-\pri{h}_p\Bigr)+4\Bigl(c_s^2\Ener\pri{\omega}+\nu(\Pres+c_s^2\Ener)\omega\Bigr)\Biggr] \, , \\
        \TP_u =& -\rho\alpha -\frac{\tau_\Heat}{4}\Biggl[\rho(h_g-h_p+4\dot{\alpha})+4c_s^2\Ener\omega\Biggr] \, , \\
        \TP_n =& \eta \Bigl[\frac{h_q}{4}-\pri{\alpha}-(\nu-2W)\alpha-\gamma\Bigr] \, , \\
        \TP_S =& \Pres h_S + c_s^2\Ener\omega -\frac{V-2\eta-\rho\tau_\Pres}{4}\Bigl(\dot{h}_g+\dot{h}_p-\pri{h}_q+4\pri{\gamma}+\nu(h_q-4\gamma)\Bigr)\\
        &-(V-\eta-\rho\tau_\Pres)\Bigl(\dot{h}_S-\frac{W}{2}(h_q-4\gamma)\Bigr)-\frac{\pri{\Ener}\tau_\Pres}{4}\Bigl(h_q-4\gamma)\, ,
    \end{aligned}
\end{equation}
for the even sector, where $\Tilde{\TP}_S=0$. We have used that in a stationary background $\mu=U=\dot{\Ener}=0$ to simplify the expressions. Additionally, we emphasize that all the terms of the stress energy tensor are written as an inviscid value (which does not involve derivatives), and a term proportional to the transport coefficients, involving first order gradients. The above expressions recover equations (76--82) in~\cite{Martin-Garcia:2000cgm} in the inviscid limit.

\section{Coordinates}\label{sec:Coordinates}

The previous expressions are valid for any coordinate chart (for the even parity case, for any stationary coordinate chart), which has the same domain of definition as the frame that we have introduced. We highlight that a frame is only locally defined and might not be defined, e.g., both in the exterior and in the interior region of a horizon, when dealing with black hole spacetimes. The issue of global definition of the frame is less important for spacetimes describing stars, which are our focus here. In the rest of this work we will fix the radial gauge, and denote the coordinates in this gauge by $\{t,r\}$. These are valid everywhere except for the irregular point $r=0$, where we need to ensure that our solutions behave in a regular manner. In these coordinates, the frame vectors are 
\begin{equation}\label{eq:Frame_To_Coordinates}
    u_Ady^A = -e^{\Phi / 2}dt \, , \qquad
    n_Ady^A = \Bigl(1-\frac{2M}{r}\Bigr)^{-1/2}dr \, ,
\end{equation}
with $M(r)$ the mass contained inside a sphere of radius $r$. The background metric is 
\begin{equation}\label{eq:Metric_In_Coordinates}
    g_{AB}dy^Ady^B = -e^{\Phi}dt^2 +\Bigl(1-\frac{2M}{r}\Bigr)^{-1} dr^2 = -e^\Phi dt^2 + e^\Lambda dr^2 \, ,
\end{equation}
where 
\begin{equation}
    \Lambda = -\log\Bigl(1-\frac{2M}{r}\Bigr) \, .
\end{equation}
At the surface of the star, $r=R_S$, the metric must match smoothly the Schwarzschild metric in the exterior. Therefore, we must have that $e^\Phi = 1-2M_S/R_S$, with $M_S$ and $R_S$ being the mass and the radius of the star, respectively.
The background scalars are 
\begin{equation}\label{eq:Scalars_To_Coordinates}
    \nu = \frac{\Phi_{,r}}{2}\sqrt{1-\frac{2M}{r}} \, , \qquad
    W = \frac{1}{r}\sqrt{1-\frac{2M}{r}} \, ,
\end{equation}
with $\mu = U = 0$. The frame derivatives, expressed in terms of coordinate derivatives, become
\begin{equation}\label{eq:Derivatives_To_Coordinates}
    \dot{X} = -e^{-\Phi/2}\frac{\partial X}{\partial t} \, , \qquad
    X^\prime = \sqrt{1-\frac{2M}{r}}\frac{\partial X}{\partial r} \, .
\end{equation}
We will also use the tortoise coordinate, defined through
\begin{equation}
    \frac{d}{dr_\star} = e^{(\Phi-\Lambda)/2}\frac{d}{dr} \, .
\end{equation}
Plugging in the background equations~\eqref{eq:Background_Einstein_General} with these coordinates yields the usual Tolman--Oppenheimer--Volkoff (TOV) equations
\begin{equation}\label{eq:TOV_Equation}
    \frac{dM}{dr} = 4\pi r^2\Ener \, , \qquad  \frac{d\Phi}{dr} =\frac{2e^\Lambda}{r^2}\Bigl(M+4\pi r^3\Pres\Bigr) \,,\qquad \frac{d\Pres}{dr} = - \frac{\Ener+\Pres}{2}\frac{d\Phi}{dr} \,  .
\end{equation}
In order to solve the TOV equations, we need to specify a particular equation of state, which for a barotrope is $\Pres=\Pres(\Ener)$, as well as a value for the central energy density. Then, the system of equation can be integrated outwards until $\Pres = 0$, which defines the surface of the star.

\section{Odd Sector}\label{sec:Odd}

We now turn our attention to the odd sector. The perturbative variables in this case are $\{k_u, k_n, \beta\}$, where $k_{u,n}$ correspond to gravitational perturbations, and $\beta$ to the fluid's angular velocity, and the only non--trivial equations are the three Einstein equations $\{\Odd_u, \Odd_n, \Odd_S\}$ and the conservation of the stress energy tensor $\CC$. We will first review the inviscid regime, and later turn our attention to including dissipative effects. The odd parity sector is studied, for a particular stellar model, in the Cowling and inverse Cowling approximations in Ref.~\cite{Boyanov:2024jge}. It is straightforward to obtain these approximations from the full viscous treatment, carried out here.

\subsection{Inviscid Regime}

In the case of a perfect fluid, the source terms~\eqref{eq:Sources_Odd} reduce to 
\begin{equation}
    \vartheta_u = -\Ener k_u -\rho\beta \, , \qquad \vartheta_n = \Pres k_n \, , \qquad \vartheta_S = 0 \, .
\end{equation}
Thus, the equation $\CC$~\eqref{eq:Odd_CSET_Projected} becomes particularly simple:
\begin{equation}\label{eq:Odd_Inviscid_Beta_Equation}
    \frac{\partial}{\partial t}\Bigl(\beta+k_l\Bigr) = 0 \, .
\end{equation}
Now, the equation $\Odd_S$ immediately yields
\begin{equation}\label{eq:Odd_kl_Inviscid}
     \frac{\partial k_l}{\partial t} = \frac{e^{(\Phi-\Lambda)/2}}{2}\Bigl(2\frac{\partial k_n}{\partial r}+\frac{d\Phi}{dr}k_n\Bigr)  \, .
\end{equation}
Finally, plugging this into $\Odd_n$ produces a decoupled wave equation for $k_n$, 
\begin{equation}
    \begin{aligned}
        -\frac{\partial^2k_n}{\partial t^2} &+ \frac{\partial^2k_n}{\partial r_\star^2}+\frac{e^{(\Phi-\Lambda)/2}}{r}\Bigl(-3+e^\Lambda(1+8\pi r^2\Pres)\Bigr)\frac{\partial k_n}{\partial r_\star} \\
        &+ \frac{e^{\Phi-\Lambda}}{4r^2}\Biggl[9-2e^\Lambda\Bigl(1+2\lambda^2-8\pi r^2(\Ener-2\Pres)\Bigr)+e^{2\Lambda}(1+8\pi r^2\Pres)^2\Biggr]k_n = 0 \, .
    \end{aligned} 
\end{equation}
Redefining $k_n = re^{-\Phi/2}\psi$, we obtain
\begin{equation}\label{eq:Master_Odd_Inviscid}
    -\frac{\partial^2\psi}{\partial t^2} + \frac{\partial^2\psi}{\partial r_\star^2}-\frac{e^{\Phi}}{r^2}\Bigl(\lambda^2-\frac{6M}{r}+4\pi r^2(\Ener-\Pres)\Bigr)\psi = 0 \, ,
\end{equation}
recovering equation (56) of~\cite{Kokkotas:1999bd}. 
The physical interpretation of this is that, as is well known, in spherical symmetry axial perturbations of stars only allow for differential rotations (the integration constant in equation~\eqref{eq:Odd_Inviscid_Beta_Equation}). This means that axial fluid modes are not oscillatory. The gravitational mode propagates according to equation~\eqref{eq:Master_Odd_Inviscid}, which becomes the usual Regge--Wheeler equation outside of the star, where $\Ener=\Pres = 0$. In particular, the master variable $\psi$ is the Regge--Wheeler function in the exterior of the star.

\subsection{Viscous Regime}

We proceed in a similar manner, but now including the viscous terms. The equation $\Odd_n$ gives rise to the following first order equation for $k_l$
\begin{equation}\label{eq:kl_Odd_Viscous}
    \frac{\partial k_l}{\partial t} = e^{-\Lambda/2}\frac{\partial}{\partial r}\Bigl(r \psi\Bigr)+16\pi\eta e^{\Phi/2}\beta \, , 
\end{equation}
and the master equation, obtained following the same steps as above, becomes
\begin{equation}\label{eq:Master_Odd_Viscous}
    \begin{aligned}
        -\frac{\partial^2\psi}{\partial t^2}&+\frac{\partial^2\psi}{\partial r_\star^2}-\frac{e^{\Phi}}{r^2}\Bigl(\lambda^2-\frac{6M}{r}+4\pi r^2(\Ener-\Pres)\Bigr)\psi \\
        =& 16\pi\eta e^{\Phi/2} \frac{\partial\psi}{\partial t} - \frac{8\pi e^{\Phi}}{r}\Bigl(2\frac{d\eta}{dr_\star}+\eta\frac{d\Phi}{dr_\star}\Bigr)\beta \, .
    \end{aligned}
\end{equation}
This equation adds two additional terms in the right hand side, both proportional to the shear viscosity: (i) a viscous damping term, proportional to $d\psi/dt$ and (ii) and explicit coupling between the rotational fluid motion (governed by $\beta$) and the master gravitational variable $\psi$. The first term would be associated to a viscous timescale 
\begin{equation}
    \tau_{\rm Diss} \sim \frac{1}{16\pi \eta} \, ,
\end{equation}
inversely proportional to the shear viscosity, consistently with~\cite{Hawking:1966qi, Press:1979rd, szekeres1971linearized}. 

The second term involves a coupling to the fluid rotational modes. In the viscous case, $\beta$ becomes a dynamical degree of freedom, satisfying the following wave equation 
\begin{equation}\label{eq:Master_Odd_Fluid_Viscous}
    \begin{aligned}
        &&-\tau_\Heat\frac{\partial^2\beta}{\partial t^2}+\frac{\eta}{\Ener+\Pres}\frac{\partial^2\beta}{\partial r_\star^2} + \eta\Bigl(\mathsf{b}_1 \frac{\partial\beta}{\partial r_\star} + \mathsf{b}_2 \frac{\partial\beta}{\partial t}+\mathsf{b}_3\beta\Bigr)\\
        &&=\mathsf{c}_1\frac{\partial^2\psi}{\partial t\partial r_\star} + \mathsf{c}_2\frac{\partial\psi}{\partial t}+\mathsf{c}_3\frac{\partial\psi}{\partial r_\star}+\mathsf{c}_4 \psi \, , 
    \end{aligned}
\end{equation}
where the coefficients $\mathsf{b}_{1,2,3}$ and $\mathsf{c}_{1,2,3,4}$ are all proportional to viscous transport coefficients, and are written in the SM. 
Therefore, axial perturbations in the presence of viscosity are described by two coupled wave equations, where both of them include explicitly dissipative terms.
An analysis of the characteristics of the problem will rapidly show two (pairs of) modes propagating with characteristic velocities $1$ and $\sqrt{\eta\tau_\Heat/(\Ener+\Pres)}$, respectively.
This shows that, in the presence of viscosity, there is a new family of modes. These modes do not have a counterpart in the perfect fluid case, since the restoring force is provided solely by shear viscosity. This is one of our main results, which highlights the need of a self-consistent treatment of dissipative effects to study relativistic oscillations of fluids. In the perfect fluid case, the axial sector describes only a propagating GW. Estimates based on energy balance~\cite{Lindblom:1983ps} can perhaps guess the dissipation of GWs through absorption by the fluid, but they fail to predict the existence of a fluid mode. Here we have shown that these fluid modes exist, and propagate with a speed characterized by the ratio between the shear viscosity $\eta$ and the dissipation timescale $\tau_\Heat$.

In the inviscid limit, i.e., when $\tau_\Heat,\eta \to 0$, the equation for $\beta$ becomes a first order equation. A straigthforward calculation shows that in that limit, we simply recover Eq.~\eqref{eq:Odd_kl_Inviscid}, when taking into account that $\partial_t \beta = \partial_t k_l$ for a perfect fluid. 

\section{Even Sector}\label{sec:Even}

Finally we study the even (or polar) perturbations. In this sector there are $7$ perturbative variables $\{h_p,h_q,h_g,h_S,\alpha,\gamma,H\}$ (we will use the enthalpy perturbation $H$, in place for the perturbation to the energy density $\omega$), and $7$ Einstein equations $\{\Even_p,\Even_q,\Even_g,\Even_u,\Even,_n,\Even_S,\Tilde{\Even}_S\}$, as well as three equations governing the conservation of the stress energy tensor. As in the previous case, we will first recover the known results for the inviscid regime. Then, we show the effects of viscosity in the Inverse Cowling Approximation (ICA). In this approximation, we freeze the fluid degrees of freedom, and show how the propagation of GWs is modified due to the absorption of GWs through dissipative effects in the fluid. Finally, we provide a set of coupled equations for the full scenario. We comment on their causal and propagation properties, and write them in a way amenible for a future numerical implementation. 

\subsection{Inviscid Regime}

One can notice immediately that the equation $\Tilde{\Even}_S$ implies that $h_g=0$, both in the inviscid and in the viscous case. Therefore we are left with only $6$ variables. Although there are approaches that reduce the problem solely to equations for the metric perturbations~\cite{Chandrasekhar:1991fi, Gundlach:1999bt}, here we write down a system of three coupled wave equations for the two metric perturbations $\{h_p,h_S\}$ and the fluid perturbation $\omega$~\cite{Allen:1997xj}. The wave character of the equations is discussed in more detail in~\ref{subsec:Characteristics}. Additionally there is a constraint which can be enforced in replacement of the corresponding wave equation, resulting in a more accurate time domain evolution~\cite{Nagar:2004ns}.

First, we can use the $\Even_n$, $\BB_n$ and $\BB_S$ equations to eliminate the time derivatives of $h_q$, $\alpha$ and $\gamma$, obtaining 
\begin{equation}\label{eq:Even_Inviscid_Nondynamical_Equations}
    \begin{aligned}
        \frac{\partial h_q}{\partial t} =& \frac{\partial h_p}{\partial r_\star}-2\frac{\partial h_S}{\partial r_\star}+h_p\frac{d\Phi}{dr_\star} \, , \\
        \frac{\partial \alpha}{\partial t} =& e^{\Phi/2}\Bigl(\frac{h_p}{4}-H\Bigr) \, , \\
        \frac{\partial \gamma}{\partial t} =& \frac{1}{2}\frac{\partial h_S}{\partial r_\star}-\frac{\partial H}{\partial r_\star}-\frac{h_p}{4}\frac{d\Phi}{dr_\star} \, .
    \end{aligned}
\end{equation}
Now, using these, we can use the combination $\Even_g-\Even_p$ to obtain the equation
\begin{equation}\label{eq:Even_HAMC_Inviscid}
    \begin{aligned}
         \frac{\partial^2h_S}{\partial r^2} = \frac{1}{4r^2}\Biggl[&2r \frac{\partial h_p}{\partial r}+2r\Bigl(r\frac{d\Lambda}{dr}-6\Bigr)\frac{\partial h_S}{\partial r}+2e^\Lambda\Bigl(\lambda^2-2\Bigr)h_S\\
         &-\frac{32\pi\rho r^2 e^\Lambda}{c_s^2}H + e^\Lambda\Bigl(\lambda^2+2-16\pi r^2\Ener\Bigr)h_p\Biggr] \, .   
    \end{aligned}
\end{equation}
Finally, the $\Even_g$, $\Even_S$ and $\BB_u$ equations lead each of them to a wave equation for the remaining variables. These are given by 
\begin{equation}\label{eq:Even_WAVE_Inviscid}
    \begin{aligned}
        -\frac{\partial^2h_S}{\partial t^2}+\frac{\partial^2h_S}{\partial r_\star^2}& = 
        -\frac{2e^{(\Phi-\Lambda)/2}}{r}\frac{\partial h_S}{\partial r_\star}+e^\Phi\Biggl[\frac{\lambda^2-2}{r^2}h_S+8\pi \rho\Bigl(1-\frac{1}{c_s^2}\Bigr)H+\frac{1}{r^2}\Bigl(e^{-\Lambda}+4\pi r^2\rho\Bigr)h_p\Biggr] \, , \\
        -\frac{\partial^2h_p}{\partial t^2} + \frac{\partial^2h_p}{\partial r_\star^2}& =
        e^{(\Phi-\Lambda)/2}\Biggl[\Bigl(\frac{2}{r}+2\frac{d\Phi}{dr}\Bigr)\frac{\partial h_p}{\partial r_\star} - \Bigl(\frac{8}{r}-4\frac{d\Phi}{dr}\Bigr)\frac{\partial h_S}{\partial r_\star}\Biggr] \\
        &+e^\Phi\Biggl[16\pi\rho \Bigl(1-\frac{1}{c_s^2}\Bigr)H + \Biggl( \frac{4e^{-\Lambda}}{r^2}-\frac{\lambda^2+4}{r^2} +8\pi(3\Ener+\Pres)+e^{-\Phi}\Bigl(\frac{d\Phi}{dr_\star}\Bigr)^2 \Biggr)h_p\Biggr]\, , \\
        -\frac{\partial^2H}{\partial t^2}+c_s^2\frac{\partial^2H}{\partial r_\star^2}& =  
        \frac{1-c_s^2}{8}\frac{d\Phi}{dr_\star}\Bigl(4\frac{\partial h_S}{\partial r_\star}-\frac{\partial h_p}{\partial r_\star}\Bigr)\\
        & +e^\Phi\Biggl[  \Biggl(\frac{c_s^2\lambda^2}{r^2}-4\pi\rho(1+3c_s^2)\Biggr)H + \Biggl(2\pi c_s^2(\Ener+3\Pres)-\frac{e^{-\Phi}}{4}\Bigl(\frac{d\Phi}{dr_\star}\Bigr)^2\Biggr) h_p        \Biggr]      \, .
    \end{aligned}
\end{equation}
These equations, together with Eq.~\eqref{eq:Even_HAMC_Inviscid}, are equivalent to Equations (9)--(11) in~\cite{Nagar:2004ns}, where the correspondence between the variables is written in Table~\ref{tab:ToNagar}.

\begin{table}[]
\begin{tabular}{|l|l|l|l|l|l|l|}
\hline
This work & $h_p$ & $h_q$ & $h_S$ & $H$ & $\Phi$ & $\Lambda$ \\ 
\hline
Ref.~\cite{Nagar:2004ns} & $2(\chi+k)$ & $2\psi$ & $k$ & $H$ & $2a$ & $2b$ \\ \hline
\end{tabular}
\caption{Comparison between the notation of this work and the notation of~\cite{Nagar:2004ns}.\label{tab:ToNagar}}
\end{table}

From these equations we can clearly see that $\{h_S, h_p\}$ encode gravitational perturbations, propagating at the local speed of light, whereas $H$ is the fluid perturbation, which propagates at the local sound speed. We could use Eq.~\eqref{eq:Even_HAMC_Inviscid} instead of the first equation in Eqs.~\eqref{eq:Even_WAVE_Inviscid} to evolve $h_S$. In that case, there would only remain one propagating degree of freedom for the gravitational sector, as expected~\cite{Nagar:2004ns}. The equations are coupled in a non--trivial way, but only first order derivatives appear as source terms. 

In the exterior, $\Ener=\Pres=H=0$, and the problem reduces to two coupled wave equations. By defining an appropriate master variable, as in~\cite{Nagar:2004ns}, this can be reduced to the Zerilli equation, which describes the propagation of GWs outside the star.

\subsection{Inverse Cowling Approximation}

In order to build the physical intuition regarding the impact of viscosity for even parity perturbations, we begin by studying the problem in the ICA~\cite{Andersson:1996ua}. This approximation is not precisely gauge invariant, and its physical information is limited~\cite{Wu:2007te}. However, it will be useful to unveil some of the strucutre that viscosity adds to the equations. We will freeze the fluid perturbations, and study only the evolution of the gravitational modes. This was used, e.g., to confirm that the $w$--modes found in the polar sector could be thought of precisely as gravitational modes~\cite{Kokkotas:1999bd}. In this case, freezing the fluid degrees of freedom in the presence of viscosity simplifies remarkably the equations, and unveils partially the role of viscosity in the propagation of gravitational degrees of freedom. After some lengthy algebra one can show that the even parity equations reduce to the following two coupled equations
\begin{equation}\label{eq:Even_ICA_Waves}
    \begin{aligned}
        -\frac{\partial^2h_S}{\partial t^2}+\frac{\partial^2h_S}{\partial r_\star^2} =& \Bigl(\dots\Bigr)_{\rm P.F.} -2\pi e^{\Phi/2}V_+ \Bigl(\frac{\partial h_p}{\partial t}+4\frac{\partial h_S}{\partial t}-\frac{\partial h_q}{\partial r_\star}\Bigr)+16\pi \eta e^{\Phi/2}\frac{\partial h_S}{\partial t}\\
        &+\frac{\pi e^{\Phi/2}}{r}\Bigl[4e^{(\Phi-\Lambda)/2}V_+ + r\Bigl(V_+-
        \frac{\rho}{c_s^2}\tau_-\Bigr)\frac{d\Phi}{dr_\star}\Bigr]h_q \, , \\
        -\frac{\partial^2h_p}{\partial t^2}+\frac{\partial^2h_p}{\partial r_\star^2} =& \Bigl(\dots\Bigr)_{\rm P.F.}-\pi V_+ e^{-\Phi/2}\Bigl(\frac{\partial h_p}{\partial t}+4\frac{\partial h_S}{\partial t}-\frac{\partial h_q}{\partial r_\star}\Bigr)+4\pi e^{-\Phi/2}\eta\frac{\partial h_p}{\partial t}\\
        &+\frac{\pi e^{-\Phi/2}}{2r}\Bigl[4e^{(\Phi-\Lambda)/2}V_+ + 8r\frac{\partial \eta}{\partial r_\star} + r\Bigl(V_+ + 4\eta - \frac{\rho}{c_s^2}\tau_-\Bigr)\frac{d\Phi}{dr_\star}\Bigr]h_q \, .
    \end{aligned}
\end{equation}
where the $(\dots)_{\rm P.F.}$ denotes the value of the right hand side of the equivalent equation in the perfect fluid case (i.e., equations~\eqref{eq:Even_WAVE_Inviscid}, setting all the fluid variables to zero). Eq.~\eqref{eq:Even_HAMC_Inviscid} is also modified, obtaining
\begin{equation}
    \begin{aligned}
        \frac{\partial^2 h_S}{\partial r^2} =& \Bigl(\dots\Bigr)_{\rm P.F.} +\pi \rho e^{\Lambda/2}\tau_\Ener\Biggl[ 2\frac{\partial h_q}{\partial r}-2e^{(\Lambda-\Phi)/2}\Bigl(\frac{\partial h_p}{\partial t}+4\frac{\partial h_S}{\partial t}\Bigr) \\
        &+\Biggl(\frac{4}{r}+\Bigl(1-\frac{1}{c_s^2}\Bigr)\frac{d\Phi}{dr}\Biggr)h_q\Biggr] \, .
    \end{aligned}
\end{equation}
Although not shown here, the equation for the first order time derivative of $h_q$ is also modified. In the modified wave equations we can see terms that are directly proportional to first order time derivatives. These are typically associated to dissipative processes (although this need not always be the case). In this case, the dissipative constant is controlled both by $V_+ = \zeta + 4\eta/3 + \rho (\tau_\Ener - \tau_\Pres)$, as well as the shear viscosity directly, $\eta$. The transport coefficient $\tau_\Heat$ does not appear altogether. The bulk viscosity only appears through $V_+$, and the remaining transport coefficients only appear through $\tau_-$. This suggests the presence of significant correlations between the transport coefficients, when looking at the perturbative relaxation of viscous NSs.

\subsection{Viscous Regime}\label{subsec:Characteristics}

Finally, we consider the full perturbative equations. In the presence of viscosity, the equations associated to the conservation of the stress energy tensor, e.g., $\BB_n,\BB_S$, become second-order equations. Indeed, one can read in~\eqref{eq:Sources_Even} that the stress energy tensor components involve first order derivatives. Therefore, we will not be able to effectively decouple $\{\alpha,\gamma\}$ by obtaining first order equations for them. In the presence of viscosity, they become dynamical degrees of freedom. This is the same as what happens for the fluid perturbations in the axial sector. 

We can still decouple $h_q$, though. In this case, we find that the modifications of the equation due to the dissipative effects are 
\begin{equation}
    \frac{\partial h_q}{\partial t} = \Bigl(\dots\Bigr)_{\rm P.F.} + 8\pi \eta \Bigl[e^{\Phi/2}\Bigl(4\gamma-h_q-\frac{8e^{-\Lambda/2}}{r}\alpha\Bigr) + 4 \frac{\partial\alpha}{\partial r_\star}+ 2\frac{d\Phi}{dr_\star}\alpha  \Bigr] \, .
\end{equation}
Then, the remaining variables $\Vec{U} = \{h_S,h_p,\alpha,\gamma,H\}$ will satisfy wave equations, which are generally coupled. Schematically, we find that the equations $\{\Even_g,\Even_S\}$ lead to second order equations for $h_S$ and $h_p$, respectively, and the conservation equations $\BB_S,\BB_l,\BB_n$ lead to second order equations for $\alpha, H, \gamma$, respectively. We still have a Hamiltonian constraint (stemming, e.g., from $\Even_p-\Even_g$), which reads 
\begin{equation}
    \begin{aligned}
        \frac{\partial^2 h_S}{\partial r^2} =& \Bigl(\dots\Bigr)_{\rm P.F.} + \pi \rho e^{\Lambda/2}\tau_\Ener \Biggl[\Biggl(\frac{4}{r}+\Bigl(1-\frac{1}{c_s^2}\Bigr)\frac{d\Phi}{dr}\Biggr)(h_q-4\gamma) \\
        &+\frac{8\lambda^2 e^{\Lambda/2}}{r^2}\alpha -2e^{(\Lambda-\Phi)/2}\Bigl(\frac{\partial h_p}{\partial t}+4\frac{\partial h_S}{\partial t}+ \frac{1}{c_s^2}\frac{\partial H}{\partial t}\Bigr)+2\frac{\partial h_q}{\partial r} - 8\frac{\partial \gamma}{\partial r}\Biggr] \, .
    \end{aligned}
\end{equation}
Interestingly, both these equations are only modified with respect to the perfect fluid case through the action of a single transport coefficient. They also recover immediately the ICA result, setting $\alpha=\gamma=H=0$. 

The remaining equations, on the other hand, are very lengthy and unilluminating.We provide the expressions for the remaining equations, as well as those for the odd parity sector, via a \texttt{Wolfram} language package~\cite{Redondo-Yuste:2024GITHUB}.
We provide these equations in a way that is ready for a numerical implementation, either in the time or in the frequency domain. The only things left to be specified are the dependence of the transport coefficients on the fundamental properties of the star, e.g., $\eta = \eta(\Ener)$, and the equation of state, $\Pres = \Pres(\Ener)$. 

Physically, we can extract much information by studying the principal part of the equation. In particular, we can understand which combinations of degrees of freedom make the system diagonal, and at which speeds do they propagate. This helps us understand which perturbative degrees of freedom are associated to which kind of modes, and can unveil potential instabilities of the formulation. Let us define $\Vec{U}$ to be a vector that contains the five evolution variables, and let $\Vec{V} = \partial_t \Vec{U}$, and $\Vec{W} = \partial_{r_\star} \Vec{U}$. Then we have the following first order system,
\begin{equation}
    \begin{aligned}
        \partial_t \Vec{U} =& \Vec{V} \, , \\
        \partial_t \Vec{W} =& \partial_{r_\star} \Vec{V} \, , \\
        \partial_t \Vec{V} =& \mathbf{A} \partial_{r_\star} \Vec{W} + \mathbf{B} \partial_{r_\star} \Vec{V} +  \mathbf{L}_1 \Vec{V} + \mathbf{L}_2 \Vec{W} + \mathbf{L}_3 \Vec{U} \, .
    \end{aligned}
\end{equation}
The operators $\mathbf{L}_i$ are lower order terms, i.e., they do not affect the principal part, gathered in the matrices $\mathbf{A}$ and $\mathbf{B}$. Therefore, setting them to zero will not change the characteristics of the problem. For a similar reason, the first equation is always lower order. The minimal system to study can be written compactly as 
\begin{equation}
    \partial_t \begin{pmatrix}
    \Vec{W} \\ \Vec{V}
    \end{pmatrix} = \begin{pmatrix}
    0 & 1_{5} \\ 
    \mathbf{A} & \mathbf{B} 
    \end{pmatrix}\partial_{r_\star}\begin{pmatrix}
    \Vec{W} \\ \Vec{V}
    \end{pmatrix}   \, ,
\end{equation}
where $1_5$ denotes the five dimensional identity matrix. The matrices $\mathbf{A}$ and $\mathbf{B}$ are given explicitly by 
\begin{equation}
    \mathbf{A} = \begin{pmatrix}
    1 & 0 & 0 & 0 & 0 \\
    0 & 1 & 64\pi\eta & 0 & 0 \\
    0 & 0 & \frac{\eta}{\rho \tau_\Heat }& 0 & 0 \\
    0 & 0 & 0 & \frac{V-\rho\tau_\Pres}{\rho\tau_\Heat} & 0 \\
    0 & 0 & -8\pi c_s^2 \eta \Bigl(1+\frac{\tau_\Heat}{\tau_\Ener}\Bigr) & 0 & -c_s^2 \frac{\tau_\Heat}{\tau_\Ener} 
    \end{pmatrix} \, , 
\end{equation}
and 
\begin{equation}
    \mathbf{B} = \begin{pmatrix}
    0 & 0 & 0 & 0 & 0 \\
    0 & 0 & 0 & 0 & 0 \\
    0 & 0 & 0 & 0 & 0 \\
    \frac{V-2\eta -\rho\tau_\Pres}{\rho\tau_\Heat} & \frac{V-\rho(\tau_\Pres-\tau_\Heat)}{4\rho\tau_\Heat} & 0 & 0 & -1-\frac{\tau_\Pres}{c_s^2\tau_\Heat} \\
    0 & 0 & 0 & -c_s^2\Bigl(1+\frac{\tau_\Heat}{\tau_\Ener}\Bigr) &  0
    \end{pmatrix} \, .
\end{equation}
The eigenvalues of the matrix, $\lambda_i$, correspond to the characteristic speeds of the problem, $c^2_i = \lambda_i^2$. These are given by 
\begin{equation}
    c^2_{\rm GW} = 1 \, \qquad c^2_{\rm Viscous} = \frac{\eta}{\rho \tau_\Heat} \, ,  \qquad c^2_{\pm} = \frac{C_1 \pm C_2}{2\rho \tau_\Ener\tau_\Heat} \, .
\end{equation}
The GW eigenvalue appears with multiplicity two, corresponding to two degrees of freedom associated to the propagation of GWs. These are associated to $\{h_S,h_p\}$, and we know that one of them is actually non--propagating, because of the additional constraint. There appears a new pair of eigenvalues, whose propagating speed is related to the ratio between shear viscosity and heat dissipation transport coefficient, similarly to the novel degree of freedom in the axial sector. We dub these modes viscous modes, and remark that they would stop being dynamical in the perfect fluid limit. As in the axial case, the restoring force for these modes, which can be associated with the rotation induced by the perturbation parameter $\alpha$, is provided by the shear viscosity. Finally, there is another pair of degrees of freedom, which involves a complicated combination of all transport coefficients, enclosed in $C_1$ and $C_2$. These are defined by 
\begin{equation}
    C_1 = \tau_\Ener\Bigl(\rho c_s^2\tau_\Heat+V\Bigr)+\rho\tau_\Pres\tau_\Heat \, , \qquad C_2=\sqrt{C_1^2 - 4c_s^2\rho\tau_\Ener\tau_\Heat^2\Bigl(\rho\tau_\Pres-V\Bigr)} \, .
\end{equation}
Out of these two degrees of freedom, one of them corresponds to enthalpy perturbations (i.e., modes associated to the $H$ equation) in the perfect fluid case. These modes get modified slightly in the presence of viscosity. The other mode has no perfect fluid counterpart. A sensible guess is that it is related to expansive motion in the fluid (i.e., to the dynamics of $\gamma$), where the bulk viscosity provides the restoring force, which makes it oscillate.

We also want to highlight that the only causality constraints emerging from these analysis (which requires all the eigenvalues to be bounded, in absolute value, by unity), are
\begin{equation}
    \begin{aligned}
         \tau_\Heat,\tau_\Ener,\tau_\Pres >& 0 \, , \\
         0 \leq \eta \leq& \rho \tau_\Heat \, , \\
         C_1^2 \geq& 4c_s^2\rho\tau_\Ener\tau_\Heat^2\Bigl(\rho\tau_\Pres-V\Bigr) \, , \\
         0\leq& C_1+C_2 \leq 2\rho \tau_\Ener\tau_\Heat \, .
    \end{aligned}
\end{equation}
These constraints are identical to the causality constraints derived, in the context of the full non-linear theory, in~\cite{Bemfica:2020zjp}. The only difference is due to the last constraint, which is not evidently equivalent to the last constraint of~\cite{Bemfica:2020zjp}, although it has a very similar structure. This shows a very non--trivial consistency check of our derivation. Notice that the constraints required for the nonlinear hydrodynamics theory to be causal are identical to the constraints of the theory linearized around any background (in particular, one that is flat)~\cite{Kovtun:2019hdm,Bemfica:2020zjp,Hoult:2024qph}.

Further exploration of these equations, both from the analytical and numerical points of view, are necessary in order to clarify the dynamical behavior of even parity perturbations of dissipative stars. However, we hope that both the analytical discussion provided, including the causal analysis, as well as the evolution equations provided in the ancillary files, can serve as a stepping stone for the community to improve our understanding of this topic.

\section{Conclusions}\label{sec:Conclusions}

The study of linearized perturbations of spherically symmetric self gravitating fluids has been crucial towards understanding the dynamical relaxation of compact stars. In this work, we have extended it to account for dissipative effects. We choose to include dissipative effects through the first order theory of BDNK, which has been shown to be causal (given some restrictions on the transport coefficients), and lead to stable evolution. 

We have employed the formalism of Gerlach and Sengupta~\cite{Gerlach:1979rw, Gerlach:1979ze, Gerlach:1980tx}, as applied by Gundlach and Mart\'in-Garc\'ia~\cite{Martin-Garcia:1998zqj, Gundlach:1999bt, Martin-Garcia:2000cgm} to reduce the perturbed Einstein equations to a set of scalar equations, in terms of gauge invariant variables. We have focused our attention on radiative multipoles, i.e., multipoles with angular index $l \geq 2$, leaving the study of the lower multipoles (including radial perturbations) to future work. 

In the axial sector, we show how the linearized system reduces to two coupled wave equations. One of them describes the modified propagation of GWs through the star, including explicitly dissipative terms. The scaling of these terms is very similar to the one proposed, e.g., in~\cite{Press:1979rd}. The second wave equation corresponds to a perturbation to the fluid velocity in the angular direction. For a perfect fluid, this does not lead to a dynamical mode. However, in the presence of viscosity, this mode becomes oscillating, propagating at a speed which depends on the local ratio between the shear viscosity and the heat dissipation rate $\tau_\Heat$. We study more thoroughly the dynamical consequences of these equations, including their matching towards the exterior solution, elsewhere~\cite{Boyanov:2024jge}. 

In the polar sector, the situation is more complicated. We have first checked that we recover the perfect fluid results in the appropriate limit. Then, we show how these are modified in the ICA, i.e., freezing the fluid degrees of freedom. In this case, there are two wave-like equations for gravitational wave modes, as well as an elliptic equation, which acts as a constraint. Remarkably, the equations are modified through the inclusion of dissipative terms, i.e., terms proportional to first order time derivatives, proportionally only to two particular combinations of the transport coefficients. This shows that although the dimensionality of the parameter space of transport coefficient is large, most of the dynamics is perhaps captured by some simple combinations of transport coefficients. 

Finally, we provide a \texttt{Wolfram} language package~\cite{Redondo-Yuste:2024GITHUB} and scripts with the perturbative equations including all dissipative effects, also in the polar sector. We studied their causal structure. We show the existence of a single propagating degree of freedom, associated to even parity GW modes, as well as three fluid modes. One of these fluid modes has the same propagation speed as the novel fluid mode in the axial sector, which we dub a shear mode. This mode does not have a perfect fluid counterpart. We argue that the last pair of modes is composed by the usual fluid mode in the even parity sector (associated to density fluctuations), whose propagation speed is modified in the presence of viscosity, and another mode which is purely viscous. Remarkably, we are able to recover the causality constraints of the BDNK theory from the analysis of the principal part of these equations, therefore checking the validity of our results. 

We did not study the effect of non radiating multipoles ($l = 0, 1$), as well as the junction conditions that match these perturbations to the perturbations in the exterior of the star (described by the Regge--Wheeler and Zerilli equations, respectively). Moreover, we simplified the problem by assuming that the thermal conductivity was negligible. These are all directions in which we plan to extend our work in the future, which also include an extension to capture the effect of magnetic fields~\cite{Armas:2022wvb}. More interestingly, the equations provided here allow us to improve our understanding of the interaction between GWs and matter. In a companion work~\cite{Boyanov:2024jge} we have shown how this occurs for axial perturbations. Studying scattering properties, as well as the spectrum of compact stars under even parity perturbations, using the equations provided in this work, is a natural continuation of this.  

\section*{Acknowledgements}

I am indebted to the developers and maintainers of \texttt{xAct}, which made possible most of this work~\cite{Martin-Garcia:2008ysv}. 

I am grateful to Vitor Cardoso, Kostas Kokkotas, and Valent\'in Boyanov for many discussions on the role of viscosity in the scattering of gravitational waves, and to Nils Andersson, Conor Dyson and David Pereñiguez for valuable comments on the manuscript. I acknowledge correspondence with Bill Press on the matter of this manuscript.

I acknowledge support by VILLUM Foundation (grant no. VIL37766) and the DNRF Chair program (grant no. DNRF162) by the Danish National Research Foundation. The Center of Gravity is a Center of Excellence funded by the Danish National Research Foundation under grant No. 184.

\section*{References}
\bibliography{References.bib}

\appendix
\clearpage 

\section{Gauge Invariance of the perturbations}

Here we show how to construct gauge-invariant perturbations. The main idea is that for an arbitrary metric perturbation $g^{(1)}_{\mu\nu}$, we will find a corresponding vector field $\EuScript{X}_\mu[g^{(1)}]$ such that the metric perturbation takes the form~\eqref{eq:Metric_SET_Decomposition} after acting with $\EuScript{X}$. 

Let us start by considering a general metric perturbation, which can be written as 
\begin{equation}\label{eq:Metric_Pert_General}
    \begin{aligned}
        g^{(1)}_{\mu\nu}dx^\mu dx^\nu =& h_{AB}\Zl dy^A dy^B + 2\Bigl(h_A\Zl_a+k_A \Xl_a\Bigr)dy^Ad\theta^a \\
        &+ \Bigl[r^2\Bigl(h_S\Zl\gamma_{ab}+\Tilde{h}_S\Zl_{ab}\Bigr)+k_S\Xl_{ab}\Bigr]d\theta^a d\theta^b \, .
    \end{aligned}
\end{equation}
Above, we omit the label $\ell$ to simplify the notation. Recall that tensor spherical harmonics are defined in the following manner~\cite{Gundlach:1999bt}:
\begin{equation}
    \begin{aligned}
        \Zl =& \mathcal{Y}^{lm} \, , \quad \Zl_a = D_a \mathcal{Y}^{lm} \, , \quad \Zl_{ab} = D_aD_b\mathcal{Y}^{lm}+\frac{\lambda^2}{2}\mathcal{Y}^{lm}\gamma_{ab} \, , \\
        \Xl_a=&\epsilon_a^b D_b\mathcal{Y}^{lm} \, , 
    \end{aligned}
\end{equation}
where we remind the reader that $D_a$ is the covariant derivative on the $2$--sphere, 
$\epsilon_a^b$ is the totally antisymmetric tensor in the two dimensional Lorentzian manifold, and $\mathcal{Y}^{lm}$ are the spherical harmonics, defined as the eigenfunctions of the equation 
\begin{equation}
    \gamma^{ab}D_aD_b\mathcal{Y}^{lm} = -\lambda^2 \mathcal{Y}^{lm} \, , 
\end{equation}
with eigenvalue $\lambda^2=l(l+1)$.
 
An arbitrary vector field can also be decomposed in terms of a $2$ dimensional vector $\xi_A$, and two scalars $\xi,\chi$, as 
\begin{equation}\label{eq:Gauge_Vector_General}
    \mathscr{X}_\mu dx^\mu = \xi_A\Zl dy^A + \Bigl(\xi \Zl_a + \chi \Xl_a\Bigr)d\theta^a \, .
\end{equation}
The gauge transformation generated by the vector $\epsilon \mathscr{X}_\mu$ (where $\epsilon$ is the small perturbative parameter) leaves the background metric invariant, and transforms the linearized metric perturbation as 
\begin{equation}\label{eq:Gauge_Transformation_Action}
    g^{(1)}_{\mu\nu} \mapsto \hat{g}^{(1)}_{\mu\nu} =  g^{(1)}_{\mu\nu} + \mathcal{L}_\mathscr{X} g^{(0)}_{\mu\nu} \, .
\end{equation}
Applying the previous decomposition, this leads to 
\begin{equation}\label{eq:Gauge_Transformation_Even}
    \begin{aligned}
        h_{AB} \mapsto& h_{AB} - 2 \nabla_{(B}\xi_{A)} \, , \qquad
        h_A \mapsto h_A -\xi_A -r^2\nabla_A \xi \, , \\
        h_S \mapsto& h_S +\lambda^2\xi-2v^A\xi_A \, , \qquad
        \Tilde{h}_S \mapsto \Tilde{h}_S -2\xi \, , 
    \end{aligned}
\end{equation}
for the even parity sector, and 
\begin{equation}\label{eq:Gauge_Transformation_Odd}
        k_{A} \mapsto k_A -r^2\nabla_A \chi \, , \qquad
        k_S \mapsto k_S -2\chi \, .
\end{equation}
From these expression it is straightforward to construct the perturbation--dependent vector field $\EuScript{X}_\mu[g^{(1)}]$, which is given by 
\begin{equation}
    \begin{aligned}
        \EuScript{X}_\mu[g^{(1)}]dx^\mu =& \Bigl(h_A-\frac{r^2}{2}\nabla_A\Tilde{h}_S\Bigr)\Zl dy^A + \frac{1}{2}\Bigl(\Tilde{h}_S\Zl_a+k_S\Xl_a\Bigr)d\theta^a \, .
    \end{aligned}
\end{equation}
Now, the metric perturbation in the gauge generated by $\EuScript{X}$, which we label as $\hat{g}^{(1)}_{\mu\nu}$ has the form 
\begin{equation}
    \hat{g}^{(1)}_{\mu\nu}dx^\mu dx^\nu = \hat{h}_{AB}\Zl dy^A dy^B + 2\hat{k}_A\Xl_ady^Ad\theta^a + r^2\hat{h}_S\Zl\gamma_{ab}d\theta^a d\theta^b \, ,
\end{equation}
which gives the form used in the main text for the metric perturbation (where for simplicity we drop the hats).

\section{Perturbations to the conservation of the stress energy tensor}

Here we write the equations describing the linearized perturbations to the conservation of the stress energy tensor, projected onto an arbitrary frame. There is only one equation in the odd sector:
\begin{equation}\label{eq:Odd_CSET_Projected}
    \begin{aligned}
        \CC_S \equiv& \dot{\vartheta}_l-\pri{\vartheta}_n+(\mu+2U)\vartheta_l-(\nu+2W)\vartheta_n+\frac{\lambda^2-2}{2r^2}\vartheta_S \\
        &=t_S(\dot{k}_l-\pri{k}_n)+k_l\Bigl(\dot{t}_S+t_S(\mu+2U)\Bigr)-k_n\Bigl(\pri{t}_S+t_S(\nu+2W)\Bigr) \, .
    \end{aligned}
\end{equation}
For the even sector the equations are given by 
\begin{equation}\label{eq:Even_CSET_Projected}
    \begin{aligned}
        \BB_l \equiv& \frac{1}{2}\Bigl(\dot{\TP}_g-\dot{\TP}_p+\pri{\TP}_q\Bigr)+U\TP_g-(\mu+U)\TP_p+(\nu+W)\TP_q-2U\TP_S - \frac{\lambda^2}{2}\TP_l \\
        & =\frac{1}{4}\Bigl[2t_g\dot{h}_g+(2t_p-t_g)\dot{h}_p-t_q\dot{h}_q+(t_p+2t_S-t_g)\dot{h}_S\Bigr] - \\
        &\frac{h_q}{4}\Bigl[2Ut_q-2(t_g-t_p)(\nu+W)+\dot{t}_q-\pri{t}_g+\pri{t}_p\Bigr] \\
        &-\frac{h_p}{4}\Bigl[2(\mu+U)t_g-2U t_p-2(\nu+W)t_q+\dot{t}_g-\dot{t}_p-\pri{t}_q\Bigr]\\
        &-\frac{h_g}{4}\Bigl[2t_p(\mu_U)-2t_q(\nu+W)-2Ut_g-\dot{t}_g+\dot{t}_p-\pri{t}_q\Bigr]-2Ut_sh_s \, , \\
        \BB_n \equiv& \frac{1}{2}\Bigl(\pri{\TP}_g+\pri{\TP}_p-\dot{\TP}_q\Bigr)+(\nu+W)\TP_p-(\mu+U)\TP_q-2W\TP_S+W\TP_g \\
        &=\frac{1}{4}\Bigl[2t_q(\dot{h}_p+\dot{h}_S)-(t_g+t_p)\dot{h}_q+(t_g+2t_p)\pri{h}_p-2t_q\pri{h}_q+t_g\pri{h}_g+2(2t_S-t_g-t_p)\pri{h}_S\Bigr]\\
        &-\frac{h_g}{4}\Bigl[2(\mu+U)t_q-2(\nu+W)t_P-2Wt_g+\dot{t}_q-\pri{t}_g-\pri{t}_p\Bigr]\\
        &+\frac{h_p}{4}\Bigl[2\Bigl(\nu t_g+(\mu+U)t_q+W(t_g+t_p)\Bigr)+\dot{t}_q+\pri{t}_g+\pri{t}_p\Bigr] \\
        &- \frac{h_q}{4}\Bigl[2\Bigl((\mu+U)(t_g+t_p)+Wt_q\Bigr)+\dot{t}_g+\dot{t}_p+\pri{t}_q\Bigr]-2Wt_Sh_S\, , \\
        \BB_s \equiv& \pri{\TP}_n-\dot{\TP}_l+(\nu+2W)\TP_n-(\mu+2U)\TP_l+\TP_S-\frac{\lambda^2-2}{2}\Tilde{\TP}_S \\
        &= \frac{1}{4}\Bigl[t_ph_p-t_qh_q+(t_g-2t_S)h_g+t_Sh_S\Bigr] \, .
    \end{aligned}
\end{equation}
%

\section{Coefficients of the odd parity equations}

In the main text, we omitted some coefficients from the master wave equations in the axial sector~\eqref{eq:Master_Odd_Viscous}--\eqref{eq:Master_Odd_Fluid_Viscous}. We write here these coefficients, which are given by:
\begin{equation}
    \begin{aligned}
        \mathsf{a} =& -\frac{e^{\Phi/2}}{r}\Biggl(\frac{\partial \log\eta}{\partial r_\star} + \frac{1}{2}\frac{d\Phi}{d r_\star}\Biggr) \, , \qquad
         \, , \\
         \mathsf{b}_1 =& \frac{4\pi r e^{(\Phi+\Lambda)/2}}{\mathcal{K}}\Biggl(2\frac{\partial \log\eta}{\partial r_\star} + \frac{d\Phi}{d r_\star}\Biggr) \, , \qquad 
        \mathsf{b}_2 = -e^{\Phi/2}\Bigl(\frac{1}{\eta}+16\pi\tau\Bigr) \, ,  \\ 
        \mathsf{b}_3 =& -\frac{4\pi e^{(\Phi-\Lambda)/2}}{r
        \mathcal{K}}\Biggl[2e^\Phi\Bigl(1+e^\Lambda(\lambda^2-1)\Bigr)+\Biggl( 4e^{(\Phi+\Lambda)/2}r-r^2 e^\Lambda \frac{d\Phi}{dr_\star}\Biggr)\frac{\partial\log\eta}{\partial r_\star}\\
        &\hspace{3cm}+re^{(\Phi+\Lambda)/2}\Bigl(\frac{d\Lambda}{d r_\star} + 3\frac{d\Phi}{d r_\star}\Bigr)\Biggr] \, , \\
        \mathsf{c}_1 =& re^{-\Phi/2}
        \Biggl(\tau - 8\pi r e^{(\Phi+\Lambda)/2}\frac{\eta}{\mathcal{K}}\Biggr) \, , \\
        \mathsf{c}_2 =& e^{-\Lambda/2}\Biggl[\tau+4\pi r \Biggl(r e^\Lambda \frac{d\Phi}{d r_\star}-6e^{(\Phi+\Lambda)/2}\Biggr)\frac{\eta}{\mathcal{K}}-8\pi r^2 e^\Lambda \frac{\eta}{\mathcal{K}}\frac{\partial\log\eta}{\partial r_\star}\Biggr] \, , \\
        \mathsf{c}_3 =& r \, , \qquad
        \mathsf{c}_4 = e^{(\Phi-\Lambda)/2} \, ,
    \end{aligned}
\end{equation}
where we have introduced for convenience 
\begin{equation}
    \mathcal{K} = \frac{d\Lambda}{d r_\star} +\frac{d\Phi}{d r_\star}= 8\pi r e^{(\Phi+\Lambda)/2}(\Ener+\Pres) \, .
\end{equation}
In the inviscid limit, the equation becomes a first order equation, given by 
\begin{equation}
    \frac{\partial \beta}{\partial t} = -e^{-\Phi/2}\Biggl(r\frac{\partial \psi}{\partial r_\star}+e^{(\Phi-\Lambda)/2}\psi\Biggr) \, ,
\end{equation}
which coincides with~\eqref{eq:Odd_Inviscid_Beta_Equation} using~\eqref{eq:Odd_kl_Inviscid} and the definition of $\psi$ in terms of $k_n$. 

\end{document}